# Inflatable Dome for Moon, Mars, Asteroids and Satellites*

Alexander A. Bolonkin
1310 Avenue R, Suite 6-F
Brooklyn, New York 11229, USA
aBolonkin@juno.com, aBolonkin@gmail.com
(718) 339-4563

**ABSTRACT**

On a planet without atmosphere, sustaining human life is very difficult and dangerous, especially during short sunlit period when low temperature prevails. To counter these environmental stresses, the author offer an innovative artificial "Evergreen" dome, an inflated hemisphere with interiors continuously providing a climate like that of Florida, Italy and Spain. The "Evergreen" dome theory is developed, substantiated by computations that show it is possible for current technology to construct and heat large enclosed volumes inexpensively. Specifically, a satisfactory result is reached by using magnetic suspended at high altitude sunlight reflectors and a special double thin film as a skin, which concentrates solar energy inside the dome while, at the same time, markedly decreasing the heat loss to exterior space. Offered design may be employed for settlements on the Moon, Mars, asteroids and satellites.

-------------------------------

## I. INTRODUCTION

The real development of outer space (permanent human life in space) requires two conditions: all-sufficient space settlement and artificial life conditions close to those prevailing currently on the Earth. (Such a goal extends what is already being attempted in the Earth-biosphere—for example at the 1$^{st}$ Advanced Architecture Contest, "Self-Sufficient Housing", sponsored by the Institute for Advanced Architecture of Catalonia, Spain, during 2006.) The first condition demands production of all main components needed for human life: food, oxidizer, and energy within the outer space and Solar System body colony. The second requisite condition is a large surface settlement having useful plants, flowers, water pool, walking and sport areas, etc. All these conditions may be realized within large 'greenhouses' [1] that will produce food, oxidizer and "the good life" conditions.

Human life in space and on other places will be more comfortable if it uses the author's macroproject proposal - staying in outer space without special spacesuit [2], p.335 (mass of current spacesuit reaches 180 kg).  The idea of this paper may be used also for control of Earth's regional and global weather and for converting Earth's desert and polar regions into edenic subtropical gardens [3]-[4].

The current conditions in Moon, Mars and Space are far from comfortable. For example, the Moon does not have useful atmosphere, the day and night continues for 14 Earth days each, there are space radiation, etc.

 Especially during wintertime, Mars could provide only a meager and uncomfortable life-style for humans, offering low temperatures, strong winds.  The distance north or south from the planet's equator is amongst the most significant measured environmental variables underlying the physical differences of the planet.

In other words, future humans living in the Moon and Mars must be more comfortable for humans to explore and properly exploit these places.

## II. 'EVERGREEN' INFLATED DOMES

Possibly the first true architectural attempt at constructing effective artificial life-support systems on the climatically harsh Moon will be the building of greenhouses. Greenhouses are maintained nearly automatically by heating, cooling, irrigation, nutrition and plant disease management equipment. Humans share commonalities in their responses to natural environmental stresses that are stimulated by night cold, day heat, absent of atmosphere, so on. Darkness everywhere inflicts the same personal visual discomfort and disorientation as cosmonauts/astronauts experience during their space-walks—that of being adrift in featureless space! With special clothing and shelters, humans can adapt successfully to the planet Mars, for example. Incontrovertibly, living in Moon, Mars is difficult, even when tempered by strong conventional protective buildings.

Our macro-engineering concept of inexpensive-to-construct-and-operate "Evergreen" inflated surface domes is supported by computations, making our macroproject speculation more than a daydream. Innovations are needed, and wanted, to realize such structures in the Moon of our unique but continuously changing life.

## III. DESCRIPTION AND INNOVATIONS

**Dome**. Our design for Moon-Mars people-housing "Evergreen" dome is presented in Fig. 1, which includes the thin inflated film dome. The innovations are listed here: (1) the construction is air-inflatable; (2) each dome is fabricated with very thin, transparent film (thickness is 0.2 to 0.4 mm) without rigid supports; (3) the enclosing film is a two-layered structural element with air between the layers to provide insulation; (4) the construction form is that of a hemisphere, or in the instance of a roadway/railway a half-tube, and part of the film has control transparency and a thin aluminum layer about 1 $\mu$ or less that functions as the gigantic collector of solar incident solar radiation (heat). Surplus heat collected may be used to generate electricity or furnish mechanical energy; and (5) the dome is equipped with sunlight controlling louvers [AKA, "jalousie", a blind or shutter having adjustable slats to regulate the passage of air and sunlight] with one side thinly coated with reflective polished aluminum of about 1 $\mu$ thickness. Real-time control of the sunlight's entrance into the dome and nighttime heat's exit is governed by the shingle-like louvers or a control transparency of the dome film.

Variant 1 of artificial inflatable Dome for Moon and Mars is shown in Fig.1. Dome has top thin double film 4 covered given area and single under ground layer 6. The space between layers 4 - 6 is about 3 meters and it is filled by air. The support cables 5 connect the top and underground layers and Dome looks as a big air-inflated beach sunbathing or swimming mattress. The Dome includes hermetic sections connected by corridors 2 and hermetic lock chambers 3. Top film has control transparency (reflectivity). That allows control temperature affecting the dome. Top film also is double. When a meteorite pushes hole in the top double film, the lower layer closes the hole and puts temporary obstacles in the way of escaping air. Dome has a fruitful soil layer, irrigation system, and cooling system 9 for support given humidity. That is a closed biosphere with closed life circle that produces an oxidizer and food for people. Simultaneously, it is the beautiful Earth place of a rest. The offered design has a minimum specific mass, about 7-12 kg/m$^2$ (air - 3 kg, film - 1 kg, soil - 3 - 8 kg). Mass of the area 10×10 m is about 1 ton.

Fig. 2 illustrates the second thin transparent dome cover we envision. The Dome has double film: semispherical layer (low pressure about 0.01 - 0.1 atm) and lower layer (high 1 atm pressure). The hemispherical inflated textile shell—technical "textiles" 1 can be woven or non-woven (films)—embodies the innovations listed: (1) the film is very thin, approximately 0.1 to 0.3 mm. A film this thin has never before been used in a major building; (2) the film has two strong nets, with a mesh of about 0.1 × 0.1 m and $a = 1 \times 1$ m, the threads are about 0.3 mm for a small mesh and about 1 mm for a big mesh.

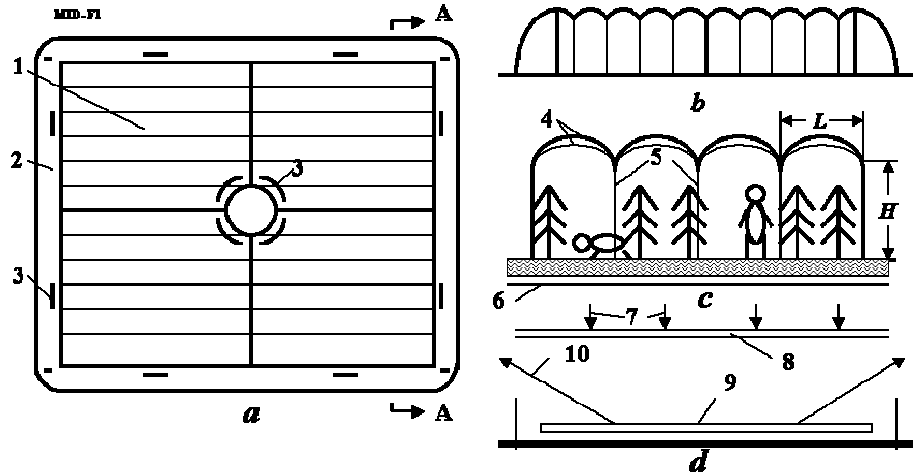

**Fig.1.** Variant 1 of artificial inflatable Dome for Moon and Mars. (a) top view of dome; (b) cross-section AA area of dome; (c) inside of the Dome; (d) Cooling system. Notations: 1 - internal section of Dome; 2 - passages; 3 - doors; 4 - transparence thin double film ("textiles") with control transparency; 5 - support cables; 6 - lower under ground film; 7 - solar light; 8 - protection film; 9 - cooling tubes; 10 - radiation of cooling tubes.

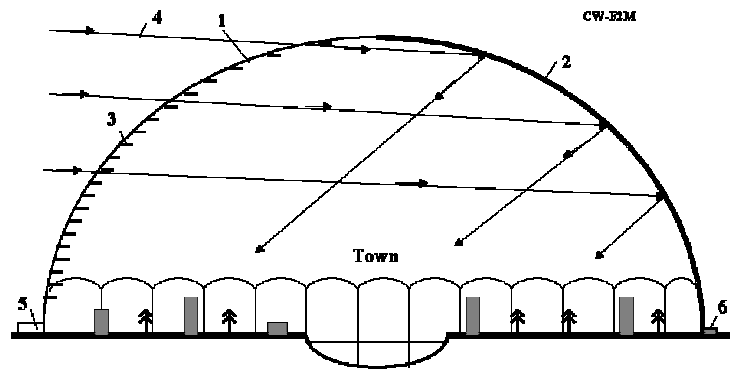

**Fig.2.** Variant 2 of artificial inflatable Dome for Moon and Mars. Notations: 1 - transparence thin double film ("textiles"); 2 - reflected cover of hemisphere; 3 - control louvers (jalousie); 4 - solar beams (light); 5 - enter (dock chamber); 6 - water extractor from air. The lower section has air pressure about 1 atm. The top section has pressure 0.01 - 0.1 atm.

The net prevents the watertight and airtight film covering from being damaged by micrometeorites; the film incorporates a tiny electrically conductive wire net with a mesh about 0.001 x 0.001 m and a line width of about 100 μ and a thickness near 1μ. The wire net can inform the "Evergreen" dome supervisors (human or automated equipment) concerning the place and size of film damage (tears, rips, punctures); the film is twin-layered with the gap — $c = 1$ m and $b = 2$ m—between the layer covering. This multi-layered covering is the main means for heat insulation and anti-puncture safety of a single layer because piercing won't cause a loss of shape since the film's second layer is unaffected by holing; the airspace in the dome's twin-layer covering can be partitioned, either hermetically or not; and part of the covering may have a very thin shiny aluminum coating that is about 1μ for reflection of non-useful or undesirable impinging solar radiation.

Offered inflatable Dome can cover a big area (town) and create a beautiful Earth conditions on a space body (fig.4a). In future, the "Evergreen" dome can cover a full planet (Moon, Mars, asteroid) (Fig.4b). Same domes can cover the Earth's regions and convert them (desert, cool regions) in beautiful gardens with controlled weather and closed life cycles.

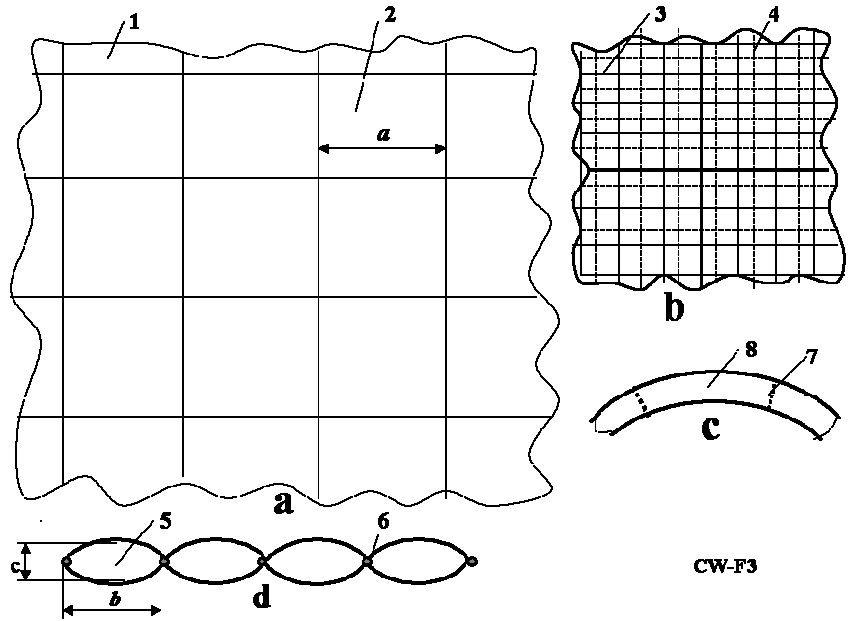

**Fig.3.** Design of "Evergreen" cover. Notations: (a) Big fragment of cover; (b) Small fragment of cover; (c) Cross-section of cover; (d) Longitudinal cross-section of cover; 1 - cover; 2 - mesh; 3 - small mesh; 4 - thin electric net; 5 - sell of cover; 6 - tubes; 7 - film partition (non hermetic); 8 - perpendicular cross-section area.

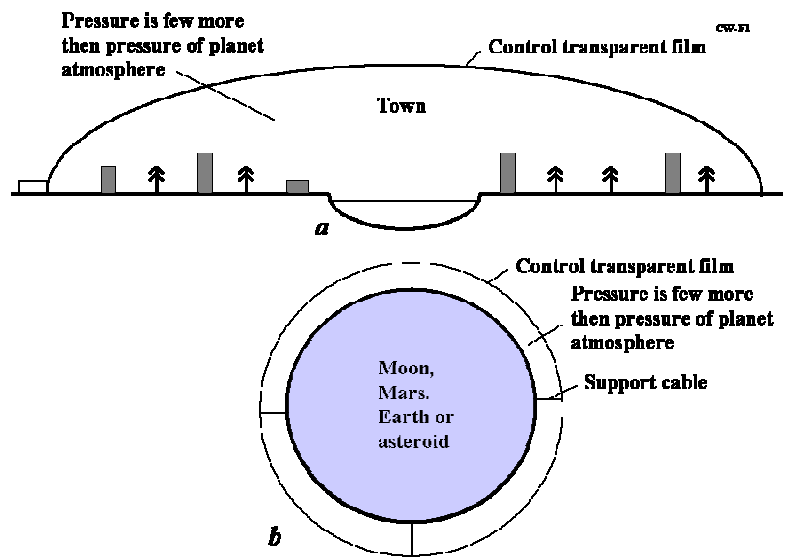

**Fig. 4.** (a) Inflatable film dome over a single town; (b) Inflatable film dome over full planet (Moon, Mars, asteroid). Same domes can cover the Earth's regions and convert them (desert, cool regions) into beautiful gardens with controlled weather and closed life cycles.

**Location, illumination and defense the human settlement from solar wind and space radiation.**

The Moon makes one revolution in about 29 Earth days. If we want to have conventional Earth artificial day and natural solar light, the settlement must locate near a Moon pole and has a magnetic control mirror suspended at high altitude in given (stationary) place (Fig. 5). For building this mirror (reflector) may use idea and theory of magnetic levitation developed by author in [5]. If reflector is made with variable focus as in [3] p. 306, fig. 16.3, then it may be as a concentrator of solar light and be used for getting energy during "night" (Earth-time).

The second important feature of the offered installation is defense of the settlement from solar wind and a cosmic radiation. It is known that the Earth's magnetic field is a defense the people from high-energy particles (protons) of a solar wind. The artificial magnetic field near Moon settlement is in hundreds times stronger than the Earth's magnetic field. It may help to defend humans. The pole location of settlement also decreases the intensity of the solar wing. Location of human settlement in polar zone Moon crater also decreases the solar wind radiation. People can move to an underground dugout, a type of bunker, at periods of high Sun activity (solar flashes, coronal mass ejections).

The theory and computation of this installation is in theoretical section. The mass of full reflector (rings, mirror, head screens is about 70 - 80 kg. If reflector is used also as powerful energy source the mass can reach 100 - 120 kg. Note, for lifting the reflector does not need a rocket. The magnetic force increases near ground (see eq. (3)). This force lifts the reflector to the needed altitude. The reflector also will be stable because it is located in magnetic hole of more powerful ground ring magnet.

The artificial magnetic field may be used also for free fly of men and vehicles, as it is described in [4]-[5]. If planet does not have enough gravity the electrostatic artificial gravity may be used [3], Ch. 15.

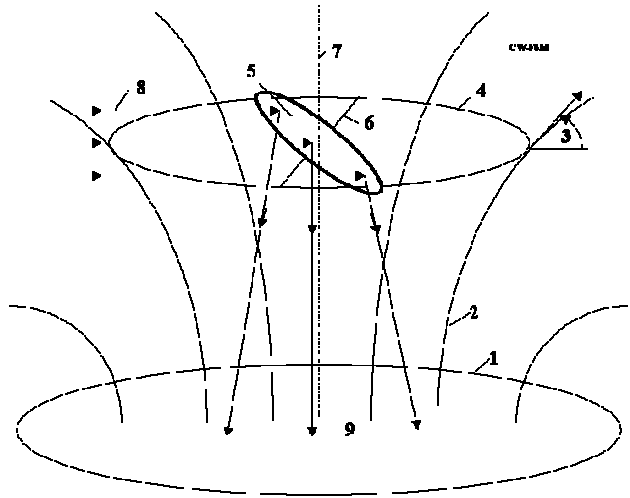

**Fig.5.** Magnetic control mirror is suspended at high altitude over human Moon settlement. Notations: 1 - superconductivity ground ring; 2 - magnetic lines of ground superconductivity ring; 3 - angle ($\alpha$) between magnetic line of the superconductivity ground ring and horizontal plate (see eq. (6)); top superconductivity ring for supporting the mirror (reflector) 5; 6 - axis of control reflector (which allows turning of mirror); 7 - vertical axis of the top superconductivity ring; 8 - solar light; 9 – human settlement. The magnetic force lifts the reflector to needed altitude.

Fig. 6 illustrates a lightweight, possibly portable house, using the same essential construction materials as the dwelling/workplace.

**Inflatable space hotel.** The offered inflatable space (satellite) hotel for tourists is shown in Fig. 7. That has the common walking area (garden) covered by a film having the controlled transparency (reflectivity), internal sections (living rooms, offices, restaurants, concert hall, storage areas, etc.).

Hotel has electrostatic artificial gravity [3], Ch.15 and magnetic field. The electrostatic artificial gravity creates usual Earth environment, the magnetic field allows people to easily fly near the outer space hotel and still be effectively defended from the solar wind.

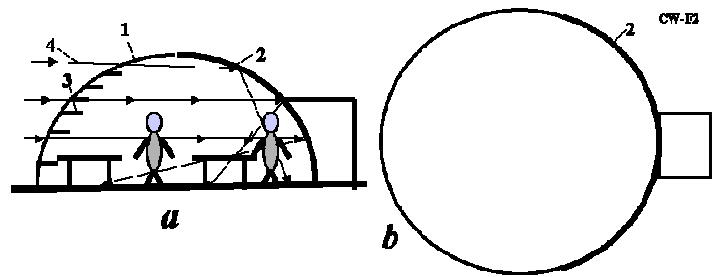

**Fig. 6.** Inflatable film house for planet. Notation: (a) Cross-section area; (b) Top view. The other notations are same with Fig. 2.

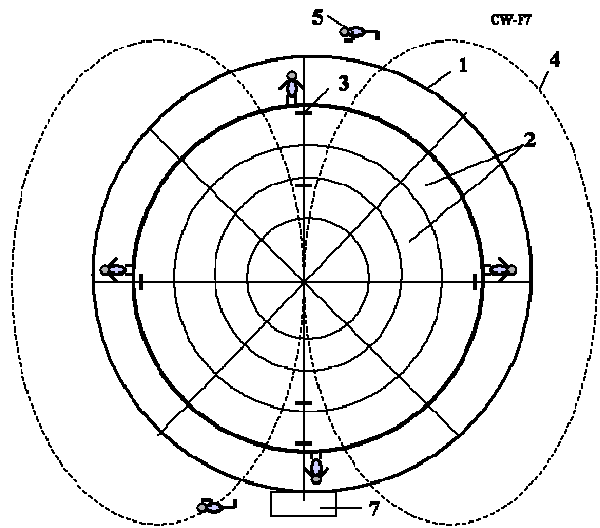

**Fig. 7**. Inflatable space (satellite) hotel. Notations: 1 - inflatable hotel (control transparency cover film); 2 - internal sections of hotel (living rooms, offices, café, music hall, storage, etc.); 3 - door and windows in internal sections; 4 - magnetic line; 5 – outer space flying person (within hotel's magnetic field, [5]); 6 - common walking area (garden). 7 - dock chamber.
   Hotel has electrostatic artificial gravity and magnetic field that will permit people to freely fly safely near the hotel even when space radiation is closely present.

   **In outer space without spacesuit.** Current spacesuit are very complex and expensive machines for living. They must, at minimum, unfailingly support human life for some period of time. However, the spacesuit makes a cosmonaut barely mobile, slow moving, prevents hard work, creates bodily discomfort, disallows to meals in outer space, has no toilet, etc. Mass of current spacesuit is about 180 kg. Cosmonauts/Astronauts—these should be combined into "Spationauts"—must have spaceship or special space home located not far from where they can undress for eating, toilet, and sleep as well as rest.
  Why do humans need the special spacesuit in outer space, or on atmosphere-less bodies of the Solar System? There is only one reason – we need an oxygen atmosphere for breathing, respiration. Human evolution in the Earth-biosphere has created the lungs that aerates the human blood with oxygen and deletes the carbonic acid. However, in a particularly harsh environment, we can do it more easily by artificial apparatus. For example, doctors, surgeons when they do surgery on heart or lugs connect their patients to the apparatus "Heart-lung machine" and temporarily stop the patient's respiration and hear-beat. In [3] p. 335 the author offered to take away part of the human blood by suture needles and to pass it through artificial "lungs".

We can design a small device that will aerate people's blood with oxygen infusion and delete the carbonic acid. If to make offshoots from main lungs arteries to this device, we would turn on (off) the artificial breathing at any time and to be in vacuum (asteroid or planet without atmosphere) or bad or poisonous atmosphere, underwater a long time. In out space we can be in conventional spacesuit defending the wearer from harmful solar light.

This idea may be checked with animal experiments in the Earth. We use the current "Heart-Lungs" apparatus and put an animal under bell glass and remove the air inside.

We can add into blood the nutrition and to be a long time without food. It is known that the humans in coma have lived many years entirely with artificial nutrition.

The life in outer space without spacesuit will be more easy, comfortable and entirely safe.

## IV. THEORY, ESTIMATION AND COMPUTATIONS OF INFLATABLE SPACE DOME

1. **Specific mass of inflatable Dome.** The mass (and relative mass) of film is summary of top double layer and support cable (Fig.1) is

$$M = \frac{pS\gamma}{\sigma}H + \frac{\pi pS\gamma}{2\sigma}L \quad or \quad \overline{M} = \frac{M}{S} = \frac{p\gamma}{\sigma}(H + \frac{\pi}{2}L), \qquad (1)$$

where $M$ is film and cable mass, kg; $p$ is air pressure, N/m$^2$; $S$ is cover area, m$^2$; $\gamma$ is specific mass of film and support cables, kg/m$^3$; $\sigma$ is safety tensile stress of film and the support cable, N/m$^2$; $H$ is height of Dome, m; $L$ is distance between support cable, m.

The needed thickness of film $\delta$ is

$$\delta = \frac{\pi p L}{2\sigma}. \qquad (2)$$

Example: Let us take $p = 10^5$ N/m$^2$; $\sigma = 10^9$ N/m$^2$ = 100 kgf/mm$^2$; $\gamma = 1800$ kg/m$^3$; $H = 3$ m; $L = 2$ m. Then $\overline{M} = 1.1 \ kg/m^2$, $\delta = 0.314 \ mm$.

2. **Magnetic stationary solar space reflector**. Magnetic intensity from ground ring

$$B \approx \mu_0 \frac{iS}{2\pi H^3}, \quad S = \pi R^2, \qquad (3)$$

where $B$ is magnetic intensity, T; $\mu_0 = 4\pi \times 10^{-7}$ is magnetic constant; $i$ is electric currency, A; $S$ is area of ground ring, m$^2$; $R$ is ground ring radius, m; $H$ is altitude of reflector, m.

*Example*: for $R = 1000$m, $H = 1000$ m; $i = 10^5$ A, magnetic intensity is $B = 6.3 \times 10^{-5}$ T.

The mass of superconductivity electric wire is

$$M_R = 2\pi R s \gamma_w, \qquad (4)$$

where $s$ is cross-section area of wire, m$^2$; $\gamma_w$ is specific mass of wire, kg/m$^3$.

For density of electric currency $j = 10^5$ A/mm$^2$ and $\gamma_w = 8000$ kg/m$^3$ the mass density of ground wire is about 50 kg. The mass of thin film heat screens which defend the wire from solar and Moon heat radiation is about 20 kg [5].

The mass of solar thin film reflector is

$$m_r = k_1 \pi r^2 \delta_r \gamma_r, \qquad (5)$$

where $r$ is reflector radius, m; $\delta_r$ is thickness of reflector film, m; $\gamma_r$ is mass density of reflector, $k_1$ is coefficient of reflector mass increasing from additional support parts (for example, from inflatable ring). For $r = 20$ m, $\delta_r = 5$ μ, $\gamma_r = 1800$ kg/m$^3$, $k_1 = 1.2$ the reflector mass is 13.6 kg.

The mass of top ring is

$$m = \frac{M}{(B j_r \cos\alpha / g_m \gamma_w) - k_2}, \qquad (6)$$

where $j_r$ is density of electric currency, A/m²; $g_m$ is gravity of planet, m/s² (for Moon $g_m$=1.62 m/s² ); $\alpha$ is angle between magnetic line and planet surface (Fig. 5); $k_2 >1$ is coefficient of top ring mass increasing from heat radiation screens. The mass of top ring is small (less 0.5 kg).

The energy emitted by a body may be computed by employment of the Josef Stefan-Ludwig Boltzmann law.

$$E = \varepsilon \sigma_s T^4, \quad [\text{W/m}^2], \qquad (7)$$

where $\varepsilon$ is coefficient of body blackness ($\varepsilon =0.03 \div 0.99$ for real bodies), $\sigma_s$ = 5.67×10⁻⁸ Stefan-Boltzmann constant. For example, the absolute blackbody ($\varepsilon= 1$) emits (at $T$ = 293 °C) the energy $E$ = 418 W/m².

The common daily average Sun heat (energy) at Earth's orbit is calculated by equation

$$Q = 86400 c q t, \qquad (8)$$

where $c$ is daily average heat flow coefficient, $c \approx 0.5$; $t$ is relative daily light time, 86400 = 24×60×60 is number of seconds in an Earth day, $q$ = 1400 W/m²s is heat flow at Earth's orbit.

The heat loss flow per 1 m² of dome film cover by convection and heat conduction is (see [6]):

$$q = k(t_1 - t_2), \quad \text{where} \quad k = \frac{1}{1/\alpha_1 + \sum_i \delta_i / \lambda_i + 1/\alpha_2}, \qquad (9)$$

where $k$ is heat transfer coefficient, $t_{1,2}$ are temperatures of the initial and final multi-layers of the heat insulators, $\alpha_{1,2}$ are convention coefficients of the initial and final multi-layers of heat insulators ($\alpha$ = 30 ÷ 100), $\delta_i$ are thickness of insulator layers; $\lambda_i$ are coefficients of heat transfer of insulator layers (see Table 1), $t_{1,2}$ are temperatures of initial and final layers, °C.

The radiation heat flow per 1 m²s of the service area computed by equations:

$$q = C_r \left[\left(\frac{T_1}{100}\right)^4 - \left(\frac{T_2}{100}\right)^4\right], \quad \text{where} \quad C_r = \frac{c_s}{1/\varepsilon_1 + 1/\varepsilon_2 - 1}, \quad c_s = 5.67, \qquad (10)$$

where $C_r$ is general radiation coefficient, $\varepsilon$ are black body rate of plates (see Table 2); $T$ is temperatures of plates, °K.

The radiation flow across a set of the heat screens is computed by equation

$$q = 0.5 \frac{C'_r}{C_r} q_r, \qquad (11)$$

where $C'_r$ is computed by equation (10) between plate and reflector.
The data of some construction materials is found in Tables 1, 2.

**Table 1**. [6], p.331. Heat Transferring.

| Material | Density, kg/m³ | Heat transfer, λ, W/m, °C | Heat capacity, kJ/kg. °C |
|---|---|---|---|
| Concrete | 2300 | 1.279 | 1.13 |
| Baked brick | 1800 | 0.758 | 0.879 |
| Ice | 920 | 2.25 | 2.26 |
| Snow | 560 | 0.465 | 2.09 |
| Glass | 2500 | 0.744 | 0.67 |
| Steel | 7900 | 45 | 0.461 |
| Air | 1.225 | 0.0244 | 1 |

As the reader will see the air layer is the best heat insulator. We do not limit its thickness $\delta$.

**Table 2**. [6], p. 465. Blackness

| Material | Blackness, $\varepsilon$ | Material | Blackness, $\varepsilon$ | Material | Blackness, $\varepsilon$ |
|---|---|---|---|---|---|
| Bright Aluminum $t = 50 \div 500\ ^oC$ | 0.04 - 0.06 | Baked brick $t = 20\ ^oC$ | 0.88 - 0.93 | Glass $t = 20 \div 100\ ^oC$ | 0.91 - 0.94 |

As the reader will notice, the shiny aluminum louver coating is excellent mean jalousie offsetting radiation losses from the dome.

The general radiation heat $Q$ computes by equation [10]. Equations [7] – [11] allow computation of the heat balance and comparison of incoming heat (gain) and outgoing heat (loss).

The heat from combusted fuel is found by equation
$$Q = c_t m / \eta, \qquad (12)$$

where $c_t$ is heat rate of fuel [J/kg]; $c_t = 40$ MJ/kg for liquid oil fuel; $m$ is fuel mass, kg; $\eta$ is efficiency of heater, $\eta = 0.5 - 0.8$.

The thickness of the dome envelope, its sheltering shell of film, is computed by formulas (from equation for tensile strength):
$$\delta_1 = \frac{Rp}{2\sigma}, \quad \delta_2 = \frac{Rp}{\sigma}, \qquad (13)$$

where $\delta_1$ is the film thickness for a spherical dome, m; $\delta_2$ is the film thickness for a cylindrical dome, m; $R$ is radius of dome, m; $p$ is additional pressure into the dome, N/m$^2$; $\sigma$ is safety tensile stress of film, N/m$^2$.

For example, compute the film thickness for dome having radius $R = 100$ m, additional air pressure $p = 0.01$ atm at top section in fig. 2 ($p = 1000$ N/m$^2$), safety tensile stress $\sigma = 50$ kg/mm$^2$ ($\sigma = 5 \times 10^8$ N/m$^2$), cylindrical dome.
$$\delta = \frac{100 \times 1000}{5 \times 10^8} = 0.0002\ m = 0.2\ mm \qquad (14)$$

The dynamic pressure from wind is (at Mars)
$$p_w = \frac{\rho V^2}{2}, \qquad (15)$$

where $\rho$ is atmospheric density, kg/m$^3$; $V$ is wind speed, m/s.

If a planet has a long nighttime, the heat protection can reduce the head losses as we can utilize inflated dome covers with more layers and more heat screens. One heat screen decreases heat losses by 2, two screens can decrease heat flow by 3 times, three by 4 times, and so on. If the inflatable domes have a multi-layer structure, the heat transfer decreases proportional to the summary thickness of its enveloping film layers.

**V. MACROPROJECTS**

The dome shelter innovations outlined here can be practically applied to many cases and climatic regimes. We suggest initial macroprojects could be small (10 m diameter) houses (Fig. 6) followed by an "Evergreen" dome covering a land area 200 × 1000 m, with irrigated vegetation, homes, open-air swimming pools, playground, concert hall.

The house and "Evergreen" dome have several innovations: magnetic suspended Sun reflector, double transparent insulating film, controllable jalousies coated with reflective aluminum (or film with control transparency) and an electronic cable mesh inherent to the film for dome safety/integrity monitoring purposes. By undertaking to construct a half-sphere house, we can acquire experience in such constructions and explore more complex constructions. By computation, a 10 m diameter home has a useful floor area of 78.5 $m^2$, airy interior volume of 262 $m^3$ covered by an envelope with an exterior area of 157 $m^2$. Its film enclosure material would have a thickness of 0.0003 m with a total mass of about 100 kg.

A city-enclosing "Evergreen" dome of 200 × 1000 m (Fig. 2, with spherical end caps) could have calculated characteristics: useful area = 2.3 × $10^5$ $m^2$, useful volume 17.8 × $10^6$ $m^3$, exterior dome area of 3.75 × $10^5$ $m^2$ comprised of a film of 0.0003 m thickness and about 200 tonnes. If the "Evergreen" dome were formed with concrete 0.25 m thick, the mass of the city-size envelope would be 200 × $10^3$ tonnes, which is a thousand times heavier. Also, just for comparison, if we made a gigantic "Evergreen" dome with stiff glass, thousands of tonnes of steel, glass would be necessary and such materials would be very costly to transport hundreds thousands, of kilometers into space to the planet where they would be assembled by highly-paid workers. Our non-woven textile (film) is flexible and plastic can be relatively cheap. The single greatest boon to "Evergreen" dome construction, whether in the Moon, Mars or elsewhere, is the protected cultivation of plants within a dome that generates energy from the available and technically harnessed sunlight.

## VI. DISCUSSION

As with any innovative macroproject proposal, the reader will naturally have many questions. We offer brief answers to the two most obvious questions our readers are likely to ponder.
(1) *Cover damage.*
The envelope contains a rip-stopping cable mesh so that the film cannot be damaged greatly. Its section, double layering structure governs the escape of air inside the living realm. Electronic signals alert supervising personnel of all rupture problems and permit a speedy repair effort by well-trained responsive emergency personnel. The top cover has a double film.
(2) *What is the design life of the film covering?*
Depending on the kind of materials used, it may be as much a decade. In all or in part, the cover can be replaced periodically.

## VII. CONCLUSION

"Evergreen" domes can foster the fuller economic development of the Moon, Mars and Earth - thus, increasing the effective area of territory dominated by humans on three celestial bodies. Normal human health can be maintained by ingestion of locally grown fresh vegetables and healthful "outdoor" exercise. "Evergreen" domes can also be used in the Tropics and Temperate Zone. Eventually, they may find application on the Moon or Mars since a vertical variant, inflatable space towers [2], are soon to become available for launching spacecraft inexpensively into orbit or interplanetary flights.

## ACKNOWLEDGEMENT

The author wishes to acknowledge R.B. Cathcart for correcting the author's English.

**REFERENCES**

(The reader find some author's works in http://Bolonkin.narod.ru/p65.htm and http://Arxiv.org   Search: Bolonkin)

[1] Bolonkin A.A., Cathcart R.B., Inflatable 'Evergreen' dome settlements for Earth's Polar Regions. Journal "Clean Technologies and Environmental Policy", Vol 9, No. 2, May 2007, pp.125-132.

[2] Bolonkin, A.A., "*Non-Rocket Space Launch and Flight*", Elsevier, London, 2006, 488 ps.

[3] Bolonkin, A.A., Control of Regional and Global Weather, 2006, http://arxiv.org search "Bolonkin".

[4] Bolonkin, A.A., "New Concepts, Ideas, and Innovations in Aerospace, Technology and Human Life". NOVA, 2007, 300 pgs.

[5] Bolonkin A.A., AB Levitation and Energy Storage, This work presented as paper AIAA-2007-4613 to 38th AIAA Plasmadynamics and Lasers Conference in conjunction with the16th International Conference on MHD Energy Conversion  on 25-27 June 2007, Miami, USA. See also http://arxiv.org  search "Bolonkin".

[6] Naschekin, V.V., *Technical thermodynamic and heat transmission*. Public House High University, Moscow. 1969 (in Russian).